\documentclass[12pt]{iopart}
\usepackage{graphicx}% Include figure files
\usepackage{dcolumn}% Align table columns on decimal point
\usepackage{bm}% bold math

\begin{document}

\title{Design, Fabrication, and Characterization of Ta$_2$O$_5$ Photonic Strip Waveguides} %Title of paper

\author{Lars Liebermeister$^1$, Niko Heinrichs$^1$, Florian B\"ohm$^1$, Peter Fischer$^{1}$, Martin Zeitlmair$^1$, Toshiyuki Tashima$^{1}$\footnote{Present address: Graduate School of Engineering
    Science, Osaka University, Toyonaka, Osaka 560-8531, Japan},  Philipp Altpeter$^1$, Harald Weinfurter$^{1,2}$, Markus Weber$^{3}$}

\address{$^1$ Fakult\"at f\"ur Physik, Ludwig-Maximilians-Universit\"at
  M\"unchen, Schellingstrasse 4, D-80799 M\"unchen, Germany}
\address{$^2$ Max-Planck-Institut f\"ur Quantenoptik, Hans-Kopfermann-Strasse
  1, D-85748 Garching bei M\"unchen, Germany}
\address{$^3$ Institut f\"ur Optik, Information und Photonik, Friedrich-Alexander University Erlangen-N\"urnberg, Staudtstrasse 7, D-91058 Erlangen, Germany}
\ead{markus.physik@gmail.com}

\date{\today}

\begin{abstract}
Efficient coupling of single quantum emitters to guided optical modes of integrated optical devices is of
high importance for applications in quantum information science as well as in the field of
sensing. Here we present the design and fabrication of a platform for on-chip
experiments based on dielectric optical single-mode waveguides (Ta$_2$O$_5$ on
SiO$_2$). The design of the waveguide is optimized for broadband (600-800 nm)
evanescent coupling to a single quantum emitter (expected efficiency: up to
36\%) and efficient off-chip coupling to single-mode optical fibers using
inverted tapers. First test samples exhibit propagation losses below 1.8 dB/mm
and off-chip coupling efficiencies exceeding 57\%. These results are promising
for efficient coupling of solid state quantum emitters
like NV- and SiV-centers to a single optical mode of a
nano-scale waveguide.
\end{abstract}


\maketitle 

\textbf{Introduction:}
On-chip evanescent coupling of light radiated by a single quantum emitter to a waveguide mode has a great potential in the field of integrated quantum information science \cite{KLM01,On-Chip Manipulation of Single Photons from a Diamond Defect}. Such devices provide e.g. an attractive platform for applications where projective on-chip measurements on photon-pairs mediate entanglement between widely separated quantum emitters \cite{Volz06,Hofmann12}. In this context, the most basic device is a single photon emitter efficiently coupled to a single optical mode \cite{Englund10,Faraon12,Liebermeister14} which then guides the photons with low loss to a detector. 

The requirements on such a photonic platform can be separated into three parts. First, the quantum system (or quantum emitter) has to be optically coupled to a waveguide. Second, the waveguide is expected to transport the photons in a confined mode with low loss. Third, the system is required to be able to couple the photons to another quantum system, device, or the macroscopic world (i.e. for instance an external photo detector). These three aspects pose different, partly even contradictory requirements on the experimental realization. Here we present the design, fabrication, and characterization of a planar single mode waveguide addressing all three aspects by featuring broadband efficient evanescent coupling to fluorescent quantum emitters combined with low-loss guidance and efficient off-chip coupling to single mode optical fibers.

\begin{figure}
\centering
\includegraphics[width=8.5cm]{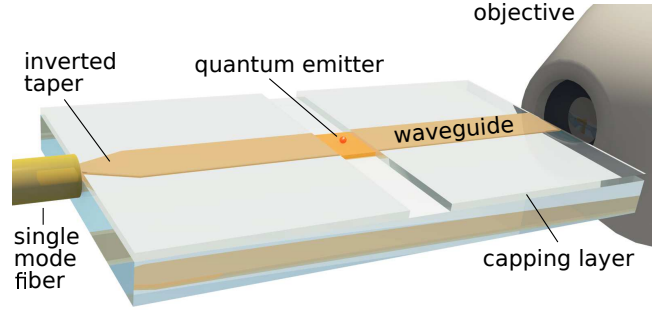}
\caption{Schematic view of a quantum emitter evanescently coupled to a planar dielectric waveguide. For efficient coupling to detectors the fluorescence photons propagate via an inverted taper and are butt-coupled into an optical single mode fiber. For optical characterization (e.g. propagation loss measurements) an additional microscope objective is used. } 
\label{fig:blender}
\end{figure}

\begin{figure}
\centering
\includegraphics[width=8.5cm]{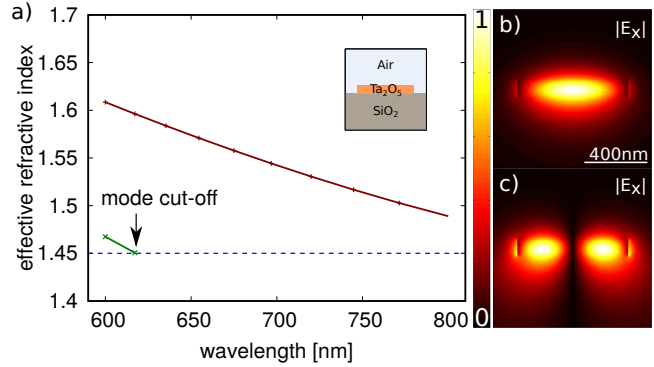}
\caption{(a) Effective refractive index of the two lowest order modes of a planar dielectric waveguide (width = 700\,nm, height = 100\,nm). Simulated mode profiles $|E_x|$ of the fundamental (b) and first order (c) guided mode at a wavelength of 600 nm.} \label{fig:n-eff}
\end{figure}

\textbf{Design:} 
Our nanoscale waveguide (see Fig. \ref{fig:blender}) is designed to provide a
versatile platform for on-chip optics allowing at the same time efficient
evanescent optical coupling to various kind of solid state quantum emitters
like colloidal quantum dots or defect centers in nanodiamonds. For the
waveguide core we have chosen the high refractive index dielectric Ta$_2$O$_5$ 
\cite{Huebner05}, whereas the lower cladding material is fused silica. 
Both materials show a low autofluorescence \cite{Hollow waveguides
with low intrinsic photoluminescence fabricated with and films}, an important
property for sensitive experiments at the single emitter level. In order to
enable efficient channeling of emitter fluorescence to the waveguide, the
vacuum electric field of the guided mode at the position of the emitter has to
be maximized \cite{Broadband waveguide QED system on a chip}. Therefore, in a
first step the evanescent electric field (at the waveguide-air interface) was
maximized by minimizing the height of the planar waveguide, yielding an
optimal value of 100\,nm. In a second step the width of the waveguide was
optimized to 700\,nm with the help of a vectorial mode solver (Mode Solutions
from Lumerical) to provide single mode operation for wavelengths from 620\,nm
to 800\,nm (see Fig. \ref{fig:n-eff}). Here we emphasize that due to the
broken top-bottom symmetry of the refractive indexes of the cladding (see
inset of Fig. \ref{fig:n-eff}(a)) the height can not be reduced below a
critical value of 90 nm to assure confinement of the fundamental mode in the
high-refractive core. 

\begin{figure}
\centering
\includegraphics[width=7cm]{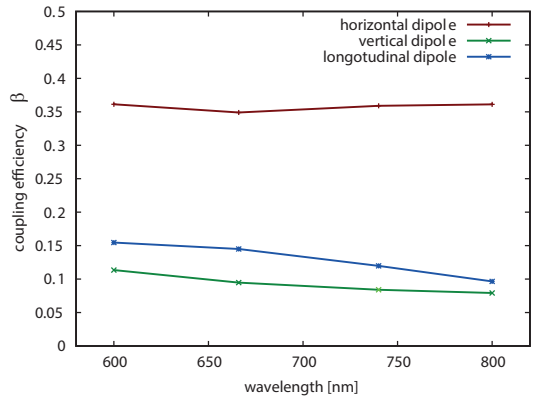}
\caption{Expected evanescent coupling efficiencies of a radiating linear dipole
placed on top of the waveguide as a function of the emission wavelength for
three different dipole orientations.} 
\label{fig:coupling}
\end{figure}

After we have fixed the waveguide geometry, we additionally performed
numerical FDTD simulations to estimate the expected radiative coupling
efficiency of a quantum emitter to the fundamental eigenmode (see
Fig. \ref{fig:n-eff}(b)). We modeled the quantum emitter as a linear dipole
positioned at the air-core interface. Similar to analytical calculations with a dipole coupled to the
nanofiber-section of a tapered optical fiber \cite{Kien04} we find broadband
coupling efficiencies (quantified by the spontaneous emission coupling
parameter $\beta$) \cite{Broadband waveguide QED system on a chip} for all three dipole polarizations (horizontal,
vertical, longitudinal) yielding typical values around (0.35, 0.1, 0.15),
respectively (see Fig. \ref{fig:coupling}).

In order to transport the guided photons to other experimental devices which
are not integrated on the chip itself off-chip-coupling to standard single
mode fibers is helpful. We use an inverted in-plane tapering of the asymmetric
waveguide \cite{Nanotaper for compact mode conversion} to adiabatically
transform the waveguide mode such that the overlap with the mode of a standard
single mode fiber is maximized.

\textbf{Fabrication:}
The waveguides and inverted tapers are fabricated using two major key
technologies, i.e. electron beam lithography (EBL) and reactive ion etching
(RIE). In more detail, nano-fabrication of the waveguide structures proceeds
as follows.

The sample wafer consists of a synthetic fused silica substrate coated with a
100\,nm thick layer of Ta$_2$O$_5$ in a sputtering process (by asphericon
GmbH, Jena, Germany). For electron beam lithography (EBL) the sample is first
spin-coated (1s at 800\,rpm and 30s at 5000\,rpm) with PMMA (950k A4 from
Microchem) as EBL-resist and then soft-baked on a hotplate for 15\,min at
170\(^\circ\)C. This results in a PMMA-layer thickness of about 190\,nm. Then,
the sample is spin-coated (30\,s at 2000\,rpm) with a conductive polymer (SX
AR-PC 5000/90.1, Allresist GmbH), which acts as discharge layer during the
electron beam exposure. Compared to a standard metal discharge layer (like
chromium) a conductive polymer promises improved roughness \cite{Nanopatterning
  of PMMA on insulating surfaces with various anticharging schemes using 30
  keV electron beam lithography} of the waveguide side-walls.

The EBL of the waveguides and inverted tapers is performed with a Raith
e\_LiNE with an applied acceleration voltage of 10\,kV. The shape of the
waveguides is written with a moving sample stage and a spatially fixed
electron beam while the inverted taper is written in the standard vector
scanning mode using an optimal dose of 95\,$\mu$C/cm$^2$. Immediately after
EBL exposure the conductive polymer is removed with clean nitrogen gas and
deionized water (LicoJET mini high pressure cleaner, Lico-Tec GmbH). The
e-beam resist is developed in 3:1 IPA:MIBK and 1.5\% MEK at room temperature
for 50\,s followed by a short cleaning-step in IPA. Now, the sample is coated
with the hard mask material (8\,nm of chromium) using an electron beam
evaporator in a high vacuum coating plant. The liftoff is mediated by
preheated DMSO at 90\(^\circ\)\,C for more than 2 hours. Finally the sample is
rinsed with acetone and IPA to remove any residual contaminants.  The hard
mask pattern is transfered to the Ta$_2$O$_5$ with inductively coupled plasma
reactive ion etching (ICP-RIE) in an Oxford Instruments Plasmalab System
100. As processing gas SF$_6$ is used diluted with Ar (4:1 in volume) with a
pressure of 5\,mtorr.  The ICP-power was optimized to 70\,W with a RF-power of
100\,W. With these parameters the 100\,nm layer of Ta$_2$O$_5$ is removed
completely within 1:40\,min. Now, the mask is removed with a liquid chromium
etchant.

\begin{figure}
\centering
\includegraphics[width=8.5cm]{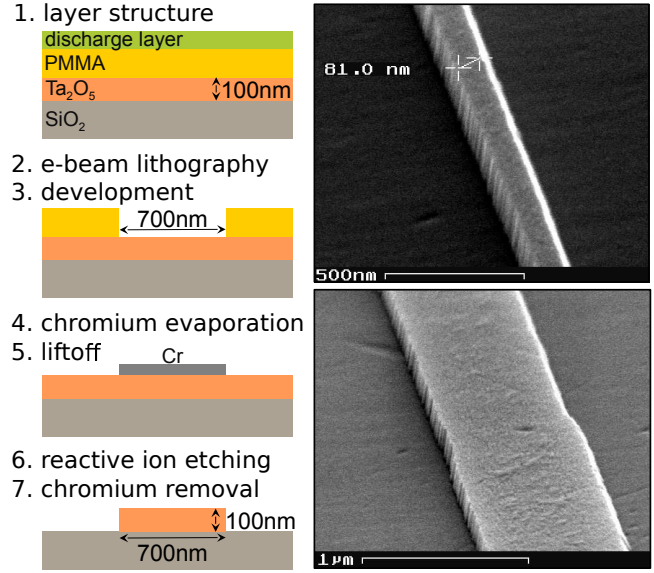}
\caption{Illustration of the main nano-fabrication steps used to structure the
  planar rib-waveguides. Scanning electron microscope (SEM) images of
  the inverted taper close to its minimum width (top, right) and transition of
  the waveguide to the inverted taper (bottom, right) after reactive ion
  etching.} \label{fig:ebl2}
\end{figure}

For butt-coupling the end facet quality of the waveguides is
crucial. Therefore a layer of 3\,$\mu$m of SiO$_2$ is sputtered on top of the
waveguides.  In order to retain an area where the waveguides are sensitive to
emitters this part is masked during the sputtering process. The end facets
were finished with a Ultratec Minipol polisher and polishing pads provided by
Ultrapol for standard FC/PC fibers. After processing a few seconds
per pad no scratches and inhomogeneities were visible in an optical inspection
microscope.

\textbf{Characterization:}
For optical characterization, the measured transmission of the fabricated
waveguides is compared for different waveguide lengths (cut-back method). On
the logarithmic scale the total transmission loss $ L_\mathrm{total}$
introduced by the waveguide is given by the sum of the in-coupling loss
$L_\mathrm{in}$, the propagation loss $ l_\mathrm{prop}$ times propagation
distance $d$, and the out-coupling loss $ L_\mathrm{out}$
\[L_\mathrm{total}(d) = L_\mathrm{in} +L_\mathrm{out} + l_\mathrm{prop}\cdot
d\]
A linear fit of the measured total transmission loss \(L_\mathrm{total}\) for
different waveguide lengths $d$ allows than to determine the propagation and
insertion loss.

\begin{figure}
\centering
\includegraphics[width=8cm]{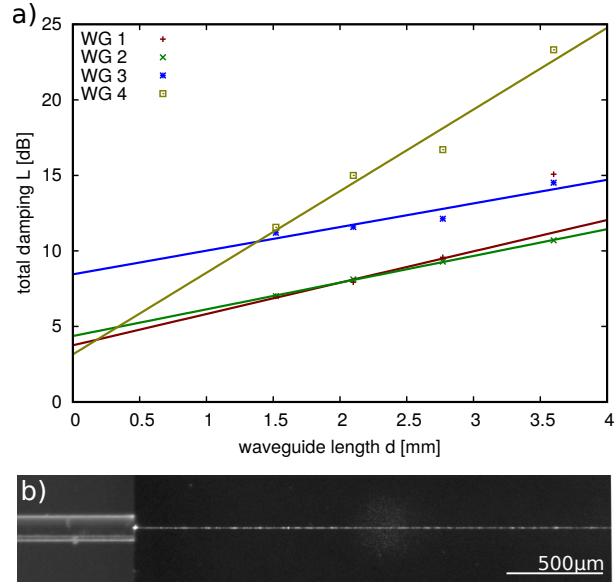}
\caption{(a) Total losses of four inspected waveguides in one sample for different sample-lengths obtained with the cut-back method. The value for waveguide 'WG 1' at 3.6\,mm has not been included into the fit as this waveguide showed a strongly scattering defect. (b) Top-view photograph of stray-light scattered by the waveguide which is butt-coupled via a single mode fiber from left.} \label{fig:probloss}
\end{figure}

The cut-back measurement is performed with a waveguide sample with an inverted
taper on one side only. The sample is shortened from the side without taper to
keep the in- and out-coupling mode geometries constant. In order to determine
the transmission, probe laser light (wavelength: 658 nm) guided by a
polarization compensated single mode fiber is coupled via butt-coupling into the waveguide (see illustration in Fig. \ref{fig:blender}). 
The light emerging from the output of the waveguide is
collected with a microscope objective and imaged onto a CCD camera or a calibrated photo-diode. Here, without inverted taper, the
high mode confinement results in a high divergence of the output mode. Within
the numerical aperture of the objective (NA=0.65) 62.3\% of this mode can be
collected (obtained from simulations of the mode-profile).  This value is the minimum achievable out-coupling loss $ L_\mathrm{out}$ in our experiment and is fixed for all measurements.  The unprocessed transmission data is plotted against the waveguide length in Fig. \ref{fig:probloss}.  From this data a propagation loss \(l_\mathrm{prop}\) below 1.8\,dB/mm as well as insertion loss $L_\mathrm{in}$ below 2.4\,dB (57\% transmission) were obtained for at least two of the 4 inspected waveguides of the chip (see Tab. \ref{tab:loss}). To avoid additional scattering losses during the cut-back measurements the sample was fully covered by a SiO$_2$ capping layer.

\begin{table}
\centering
\label{tab:loss}
 \begin{tabular}{r|r|r} 
 \label{tab:loss}
  Waveguide & Prop. loss & Insertion loss  \\
 \hline
   1&(2.07$\pm$0.24)\,dB/mm&(1.69$\pm$0.53)\,dB\\
 2&(1.77$\pm$0.03)\,dB/mm&(2.31$\pm$0.08)\,dB\\
 3&(1.56$\pm$0.40)\,dB/mm&(6.39$\pm$1.05)\,dB\\
 4&(5.40$\pm$0.76)\,dB/mm&(1.09$\pm$1.99)\,dB\\
 \hline
\hline
\end{tabular}
\label{tab:loss}
\caption{Table with propagation loss and insertion loss for 4 representative waveguides on the same sample obtained from the slope and the y-axis intercept of the fit to the total transmission for 4 different lengths using the cut-back method. The values of the insertion loss are obtained by subtracting the simulated value for the out-coupling loss introduced by the limited NA of the collection optics  ($ L_\mathrm{out} = 0.632$) from the y-axis intercept provided by the linear fit. The error-ranges originate from the least square fit asymptotic standard errors.} 
\label{tab:loss}
\end{table}

\textbf{Discussion:} A dielectric on-chip waveguide has been designed to provide a versatile platform for on-chip
experiments in quantum information science and sensing applications. The waveguide promises
moderate coupling efficiencies of about 35\% to a dipole placed on its surface and aligned parallel
to the long axis of the waveguide. To realize first samples, a fabrication process
has been developed based on nano-fabrication techniques such as electron beam lithography
and reactive ion etching. The electron beam lithography process allows to pattern continuous
waveguides of several millimeters in length which can be connected to arbitrary structures
without stitching errors. The overall process was optimized to get reproducible propagation losses below $1.8$ dB/mm at an operation wavelength of 658 nm. This experimental finding is in good agreement with an estimated value of $1.8$ dB/mm on basis of the Payne-Lacey model, which uses the measured sidewall-roughness (determined via independent AFM investigations \cite{Heinrichs2015}) as input parameter. In the literature propagation losses as low as $0.4$ dB/cm have been reported \cite{Fabrication and optimization of Tantalum Pentoxide waveguides for optical micro-propulsion} using the same material system (Ta\(_2\)O\(_5\) on SiO\(_2\)), but with much bigger dimensions (200 nm x 2...10 $\mu$m) and light with a longer wavelength (1070 nm). However, with a system more comparable in geometry, Fu et al. \cite{Coupling of nitrogen-vacancy centers in diamond to a GaP waveguide} found propagation loss of 4.6\(\pm\)3 dB/mm for a rig waveguide of 1 $\mu$m in width and 160 nm in hight (GaP on diamond).

To provide efficient off-chip coupling to standard single-mode fibers an inverted taper was added to the waveguide. For our fabricated samples we find typical off-chip coupling efficiencies of 57\%. With the values achieved for propagation length and off-chip coupling, combined with the expected coupling-efficiency from the emitter to the waveguide, a total coupling from the emitter
to a standard fiber-coupled APD-detector of above 10\% is realistic. Further progress in fabrication can then pave the way to slot waveguides \cite{Broadband waveguide QED system on a chip}, cavities with distributed Bragg reflectors, or even a combination of both \cite{Silicon photonic slot waveguide Bragg gratings and resonators} promising emitter to waveguide coupling of up to 65\% \cite{Fischer2015}. 

\section*{Acknowledgments}
We acknowledge funding from the DFG through the excellence cluster NIM and the
Forschergruppe 1493. TT acknowledges support from the Japanese Society for the
Promotion of Science.


\end{document}